Łukasz Górski
Interdisciplinary Centre for Mathematical and
Computational Modelling
University of Warsaw
Faculty of Mathematics and Computer Science
Nicolaus Copernicus University in Toruń

# AGAINST EXPLAINABLE ARTIFICIAL INTELLIGENCE IN LAW: WHY JUSTIFIABLE AI MATTERS. A CREDIT SCORING EXAMPLE

**Abstract**

Artificial intelligence-based solutions offer new efficiency-increasing possibilities in many applications, including credit scoring. Yet, the increasing sophistication of machine-learning models in use raises concerns regarding many of their aspects, explainability notwithstanding. We review the relevant EU legal background and integrate this review with insights from technical sciences to interpret relevant legal provisions in the light of technological possibilities. We reject the narrow interpretations of the right to explanation and suggest the broad one, which encompasses not only a technical explanations but also a legal justification as the only one that allows to safeguard the creditors rights in an operative manner.

## I. INTRODUCTION

Artificial intelligence (AI) systems are transforming the economies, including the financial sector and how creditworthiness is assessed. They now underpin decision-making processes once reserved for human judgment. As these systems grow more opaque, the ability to explain how they arrive at particular outcomes becomes both a technical and legal necessity – especially in domains where such decisions carry serious consequences for individual rights.

In response to this growing opacity, the concept of explainable AI (XAI) has emerged as a central pillar of responsible AI deployment[1]. Various legislative instruments, including the EU's Consumer Credit Directive[2], the AI Act[3], and to some extent GDPR[4], incorporate the right to explanation as a mechanism for enhancing transparency and accountability in diverse contexts. Yet, these rights are often grounded in the technical forms of explanation generation. (e.g. a general information that a 'black box' AI system was involved), or too much – they are presented with concrete mathematical formulas and weights (parameters) that effected in a given result, without describing how exactly that connects to their situation[5]. CJEU has also acknowledged that use of AI systems can hinder transparency and present addressees with decisions whereas concrete reasoning is impossible to trace, thus depriving them of rights enshrined in art. 47 of the Charter of Fundamental Rights[6].

In this paper, drawing on doctrinal legal analysis, comparative perspectives, and practical experience with the development of explainable systems, we argue that technical explanations, even when made available, frequently fail to deliver actionable knowledge to end users. Instead of empowering consumers, they often obscure the rationale behind decisions in layers of complexity, thereby creating a form of regulatory fiction: a formal compliance with transparency requirements that lacks substantive effect.

---

[1] Alejandro Barredo Arrieta and others, 'Explainable Artificial Intelligence (XAI): Concepts, Taxonomies, Opportunities and Challenges toward Responsible AI' (2020) 58 Information Fusion 82.

[2] Directive (EU) 2023/2225 of the European Parliament and of the Council of 18 October 2023 on credit agreements for consumers and repealing Directive 2008/48/EC [2023] OJ L, L2023/2225, 1; hereinafter CCD.

[3] Regulation (EU) 2024/1689 of the European Parliament and of the Council of 13 June 2024 laying down harmonised rules on artificial intelligence and amending Regulations (EC) No 300/2008, (EU) No 167/2013, (EU) No 168/2013, (EU) 2018/858, (EU) 2018/1139 and (EU) 2019/2144 and Directives 2014/90/EU, (EU) 2016/797 and (EU) 2020/1828 (Artificial Intelligence Act) [2024] OJ L L2024/1689, 1; hereinafter AIA.

[4] Regulation (EU) 2016/679 of the European Parliament and of the Council of 27 April 2016 on the protection of natural persons with regard to the processing of personal data and on the free movement of such data, and repealing Directive 95/46/EC (General Data Protection Regulation) [2016] OJ L L2016/119, 1; hereinafter 'the GDPR'.

[5] Madalina Busuioc, Deirdre Curtin and Marco Almada, 'Reclaiming Transparency: Contesting the Logics of Secrecy within the AI Act' (2023) 2 European Law Open 79, 87.

[6] Case C-817/19 Ligue des droits humains v Conseil des ministres [2022] ECLI:EU:C:2022:491; Ljubisa Metikos and Jef Ausloos, 'The Right to an Explanation in Practice: Insights from Case Law for the GDPR and the AI Act' (Social Science Research Network, 6 August 2024) 20 <https://papers.ssrn.com/abstract=4996173> accessed 24 October 2024.

We hypothesize the concept of justifiability is a superior standard. We build on earlier philosophical results regarding the process of scientific discovery and judicial decision-making to show their applicability in the era of AI. Whereas explainability focuses on the internal logic of a system (the context of discovery), justifiability emphasizes external reasoning (the context of justification), grounded in legal, procedural, and social norms. We argue that in B2C (business-to-consumer) contexts like credit scoring, only justifiable AI can meaningfully fulfil the legal promise of the right to explanation. Credit scoring was chosen as an illustrative examples for our claims, as the financial institutions are at the forefront of AI adoption[7]. This also allows paper to remain focused. The sole notion of using AI for credit scoring has already been acknowledged on the basis of the Polish law[8] (with various other applications considered[9]), and this paper expands the aspect of its explainability.

To validate this paper's claim, we examine three central research questions (RQs): RQ1: What limitations do current technical XAI solutions present in the context of credit scoring? RQ2: How should the right to explanation be interpreted in light of these limitations to ensure legal adequacy? RQ3: Can a concept of justifiability be introduced as a superior standard?

"Justifiable AI" is novel term, still scarcely present in the literature. It is connected with retrieval of legal rationale or supporting evidence with the use of AI chatbots[10]; the same need for presentation of rationale and possible mismatch was also mentioned in the context of tax law[11]. In this paper, we put this concept of "justifiable AI" into concrete application of financial scoring and show the inadequacy of explanation generation methods together with the need of broad interpretation of the term "explanation". According to our knowledge, this is the first time such exercise is performed, especially in the light of the fact that legal and technical environment changes dynamically – for example many papers on explainability do not acknowledge the existence of art. 86 of the AIA, as it was added late in the drafting process[12], even though it established a "right to explanation", with an obligation to provide "clear and meaningful explanation of the role of the AI system in the decision-making procedure and the main elements of the decision taken". This, however, applies only to decisions with legal or similarly significant effects that have an adverse impact on their health, safety or fundamental rights. Art. 86 furthers the general aim of the AIA, with its various provisions aimed to protect fundamental rights, health and safety by making elements of the system auditable. This focus on the explainability being not an aim in itself, but rather an instrument to further the well-being of EU's citizens is also present in the AIA's recitals. Recital 27 notes that transparency is composed, *inter alia*, of explainability, and – in general – transparency itself is one of the underpinning ethical principles that make the deployment of AI trustworthy and ethically sound (recital 27). Recital 59 notes how the right to defence, as well as the presumption of innocence need to be coupled with explainability for them to be protected. Credit scoring systems are deemed to be high-risk in the context of AIA, "since they determine those persons' access to financial resources or essential services such as housing, electricity, and telecommunication services" (recital 56 of the CCD).

---

[7] Antonella Sciarrone Alibrandi, Maddalena Rabitti and Giulia Schneider, 'The European AI Act's Impact on Financial Markets: From Governance to Co-Regulation' 10.

[8] Antonina Chłopecka, 'The Use of Artificial Intelligence in Assessing a Bank Customer's Debt Capacity' [2023] Studia Iuridica 73.

[9] Tomasz Szanciło, 'Artificial Intelligence and Case Categories in Civil Proceedings' [2024] Studia Iuridica 7.

[10] Sabine Wehnert, 'Justifiable Artificial Intelligence: Engineering Large Language Models for Legal Applications' (arXiv, 27 November 2023) <http://arxiv.org/abs/2311.15716> accessed 27 June 2025.

[11] Marco Almada and others, 'Towards eXplainable Artificial Intelligence (XAI) in Tax Law: The Need for a Minimum Legal Standard' (2022) 14 World tax journal.

In this paper, methodologically, we adopt an interdisciplinary approach. We combine doctrinal analysis of EU law with comparative references and insights drawn from our own system development work and user feedback. The structure of the paper is as follows: after introductory remaks, Chapter II outlines technical explainability methods in credit scoring. Chapter III explores the legal frameworks governing explanation rights. Chapter IV presents our normative argument for a shift toward justifiability. The final section offers conclusions.

## II. AI AND XAI SOLUTIONS FOR CREDIT SCORING

AI-based systems have already found their place in a number of solutions employed on the financial markets. Those include: fraud detection, financial analysis, or risk management[13]. FinTech systems for credit scoring involve the use of AI methods based on

[12] The full provision states that "

1. Any affected person subject to a decision which is taken by the deployer on the basis of the output from a high-risk AI system listed in Annex III, with the exception of systems listed under point 2 thereof, and which produces legal effects or similarly significantly affects that person in a way that they consider to have an adverse impact on their health, safety or fundamental rights shall have the right to obtain from the deployer clear and meaningful explanations of the role of the AI system in the decision-making procedure and the main elements of the decision taken.

2. Paragraph 1 shall not apply to the use of AI systems for which exceptions from, or restrictions to, the obligation under that paragraph follow from Union or national law in compliance with Union law.

3. This Article shall apply only to the extent that the right referred to in paragraph 1 is not otherwise provided for under Union law."

[13] Akoh Atadoga and others, 'The Intersection Of Ai And Quantum Computing In Financial Markets: A Critical Review' (2024) 5 Computer Science & IT Research Journal 461, 462.

various sources, including historical data, particular borrowing habits, or projected cash flows[14]. Use of algorithmic methods allows to speed up scoring time effecting in lowering the expenditures of the financial institutions, and – at the same time – facilitates credit action broadening[15]. Methodologically similar concept, social scoring, is explicitly prohibited by the AIA[16]. Whilst the scoring procedures for credit granting involve the analysis of financial flows, prohibited social scoring involves insights based on multifaceted and often cross-contextual personal or behavioural data, *prima facie* not concerned with the specific decision at hand. This can lead to detrimental or unfair treatment that is detached from the original purpose for which the data was collected. Yet, even in the context of credit scoring, there is a sizable body of contemporary research that also involves introduction of algorithms that, in addition to financial data, use e.g. social networks analysis, or other similar background checks that potentially allow to discover the social and personal background of a given person[17]. Seemingly neutral data points may thus lead to the data on e.g. race or gender to be eventually involved in the scoring process and proxied so that those correlations escape systems' analysis[18]. To alleviate this danger, the CCD explicitly states that "[t]he assessment of creditworthiness shall be carried out on the basis of relevant and accurate information on the consumer's income and expenses and other financial and economic circumstances which is necessary and proportionate to the nature, duration, value and risks of the credit for the consumer." (art. 18(3)).

From the technical side, there have been many algorithms that have been tried for credit scoring with the use of financial data. Whilst designing algorithms for credit scoring, financial institutions often focus on those that are simple and thus easy for their personnel to

[14] Anil Kumar, Suneel Sharma and Mehregan Mahdavi, 'Machine Learning (ML) Technologies for Digital Credit Scoring in Rural Finance: A Literature Review' (2021) 9 Risks 192, 2.

[15] ibid.

[16] Art. 5(1)(c) prohibits "

(c) the placing on the market, the putting into service or the use of AI systems for the evaluation or classification of natural persons or groups of persons over a certain period of time based on their social behaviour or known, inferred or predicted personal or personality characteristics, with the social score leading to either or both of the following:

(i) detrimental or unfavourable treatment of certain natural persons or groups of persons in social contexts that are unrelated to the contexts in which the data was originally generated or collected;

(ii) detrimental or unfavourable treatment of certain natural persons or groups of persons that is unjustified or disproportionate to their social behaviour or its gravity;"

[17] Yanhao Wei and others, 'Credit Scoring with Social Network Data' (2016) 35 Marketing Science 234.

[18] Mikella Hurley and Julius Adebayo, 'CREDIT SCORING IN THE ERA OF BIG DATA' (2016) 18 Big Data.

work with[19], various forms of sophisticated techniques were tested by researchers, including the ones focused on the explainability (other works can be consulted for details that are omitted here for brevity[20]).

Broadly speaking, explainable AI (XAI) is an umbrella term that denotes a range of methods that can be used to create AI systems that in their working are understandable for humans[21]. How this aim is attained depends on the exact ML (machine-learning) model used. Herein, *interpretable* and *explainable* models can be distinguished. The former term applies to categories of systems that can in principle be understood by a person sufficiently well-versed in the intricacies of ML. This includes methods such as logistic regression, whereas the way it transforms credit scoring data (like numerically expressed age, yearly income, or personal liabilities) is described by mathematical formulae. Whilst some expertise is needed, such models are inherently understandable[22].

In the case of non-interpretable models, often referred to as "black boxes," external methods are typically employed to enhance explainability. For instance, modern neural networks are generally large and complex, comprising numerous layers and nonlinear transformations of input data, which render their internal workings difficult to follow. In their case external methods can be used to trace their working. Commonly used solutions are SHAP, or Grad-CAM[23]. When approached functionally, three categories of such solutions can be distinguished, though alternative typologies are possible. *Feature importance-based* explanations connect the particular input data with a weight this data has when arriving at the final result (e.g. age of 55 years contributed in 40% towards the rejection). One popular algorithm in this category is LIME, a *surrogate-based* explainer. This means it builds a simpler interpretable model – a surrogate – around the usually more complex one that is explained. *Counterfactual* explainer show how the input data should be changed so that the model's decision is reversed (e.g. should you be of 50 years or younger, the credit application would have been approved). In the case of *actionable recourse* explainer, the explainee is instructed what action they should take to reverse the result (e.g. a person might find it very difficult to become younger, but might find a recourse in cancelling their credit cards[24]). The explanations provide understanding (e.g. by uncovering the relationship between model's inputs and outputs), contestability (e.g. provide the explainee with knowledge the model discriminates by age, something that can be used as an argument in subsequent legal proceedings), and recourse (actionable knowledge)[25]. Direct actionability is especially concerned with *local* explanations (applicable for a given case) and those can be contrasted with *global* ones – they give a general overview of the system's working, without reference to a concrete case (e.g. what weight is assigned to the yearly income by the system in the totality of cases?)[26].

---

[19] Ouren Kuiper and others, 'Exploring Explainable AI in the Financial Sector: Perspectives of Banks and Supervisory Authorities' in Luis A Leiva and others (eds), *Artificial Intelligence and Machine Learning*, vol 1530 (Springer International Publishing 2022) 6 <https://link.springer.com/10.1007/978-3-030-93842-0_6> accessed 19 December 2024.

[20] Jurgita Černevičienė and Audrius Kabašinskas, 'Explainable Artificial Intelligence (XAI) in Finance: A Systematic Literature Review' (2024) 57 Artificial Intelligence Review 216.

[21] Łukasz Górski and others, 'Exploring Explainable AI in the Tax Domain' [2024] Artificial Intelligence and Law 3.

[22] ibid 24.

[23] Łukasz Górski and Shashishekar Ramakrishna, 'Explainable Artificial Intelligence, Lawyer's Perspective', *Proceedings of the Eighteenth International Conference on Artificial Intelligence and Law* (Association for Computing Machinery 2021) <https://doi.org/10.1145/3462757.3466145>.

In general, when financial institutions are surveyed, explainable AI scores highly on their priority lists for implementation and tends to be integrated within the internal procedures and ethical guidelines[27]. Explainability is at same time identified as one of the most important facet as well as the challenge of AI[28]. Whilst the right to explanation can be interpreted from the general ethical principles[29], the positive law is a source that is one of the main factors for introducing internal rules[30]. In the absence of factors stemming from positive law, the motivation for organization to implement appropriate rules is significantly diminished. For example, in the case of AI systems used in the legal environment, at least one of the survey indicated the logic of effectiveness prevails over the need for an explanation[31]. In other words, with the lack of relevant legal encouragements, it suffices for AI-based tools to offer solutions that are efficacious and not necessarily explainable.

---

[24] Timo Freiesleben and Gunnar König, 'Dear XAI Community, We Need to Talk! Fundamental Misconceptions in Current XAI Research', *World Conference on Explainable Artificial Intelligence* (Springer 2023) 5.

[25] ibid.

[26] Christoph Molnar, *Interpretable Machine Learning* (Lulu.com 2020) <https://christophm.github.io/interpretable-ml-book/index.html#summary> accessed 16 May 2022.

[27] Kuiper and others (n 19) 9.

[28] Martin Ebers, 'Regulating Explainable AI in the European Union. An Overview of the Current Legal Framework (s)' [2020] An Overview of the Current Legal Framework (s)(August 9, 2021). Liane Colonna/Stanley Greenstein (eds.), Nordic Yearbook of Law and Informatics 3.

[29] Fleur Jongepier and Esther Keymolen, 'Explanation and Agency: Exploring the Normative-Epistemic Landscape of the "Right to Explanation"' (2022) 24 Ethics and Information Technology 49, 5–6.

[30] Kuiper and others (n 19) 2.

[31] Michał Jackowski and others, 'First Global Report on the State of Artificial Intelligence in Legal Practice' (Liquid Legal Institute 2023) 35.

## III. LEGAL REQUIREMENTS FOR EXPLAINABLE ARTIFICIAL INTELLIGENCE FOR CREDIT SCORING IN THE EU LAW

There has been a rising awareness regarding the need for regulating the AI-based solutions, credit scoring notwithstanding. Around the world, this has been approached with various degree of comprehensiveness, implemented either as vertical or horizontal regulation. European Union has chosen a comprehensive path, with the AIA being a flagship project for AI regulation[32]. However, as far as the this paper's credit scoring subject-matter is concerned, it is of limited application[33]. Firstly, due to a comprehensive preexisting technology-neutral regulatory frameworks already applying to financial institutions[34], it lists several exemptions and special provisions – cf. *inter alia* its arts. 17(4), 18(3), 26(5). Secondly, this paper will also forgo the AIA's arts. 13 and 14, as they relate to transparency obligations between system's provider (developer) and deployer (operator). Their explainability-related provisions mainly have to do with the deployer's being provided with the necessary tools and knowledge to interpret system's output and do not directly pertain to B2C relationship. As far as the B2C relationship is concerned, even though the AIA grants the right to an explanation in its art. 86, it is also not applicable in the case of credit scoring.

Explicit derogation (art. 86(3) AIA) states that this provision applies only insofar as the right is not provided by other EU laws[35]. The provision central for XAI vis-á-vis credit scoring is thus art. 18(8) of the CCD. It ensures that consumers are provided with a clear and comprehensible explanation of the assessment of creditworthiness, including the *logic* and risks involved in the automated processing of personal data as well as its *significance* and *effects* on the decision, should the creditworthiness assessment involve the use of automated personal data processing (art. 18(8)(a), emphasis added)[36].

---

[32] Daniel Perez Del Prado, 'The Challenges of Algorithm Management: The Spanish Perspective' (2024) 29 Białostockie Studia Prawnicze 131, 134.

[33] Other sources can be referred for its critical overview, cf. e.g. Sandra Wachter, 'Limitations and Loopholes in the EU AI Act and AI Liability Directives: What This Means for the European Union, the United States, and Beyond' [2024] SSRN Electronic Journal <https://www.ssrn.com/abstract=4924553> accessed 14 November 2024.

[34] Deutsche Bundesbank, 'The Use of Artificial Intelligence and Machine Learning in the Financial Sector' [2020] URL: https://www. bundesbank. de/resource/blob/598256/d7d26167bceb18ee7c0c296902e42162/mL/2020-11-policy-dp-aiml-data. pdf.

[35] Žiga Škorjanc, 'The Right to Explanation of a Credit Score: A Holistic Approach under the GDPR, AI Act, and Directive (EU) 2023/2225 on Credit Agreements for Consumers' (2025) 6 Global Privacy Law Review 104 <https://kluwerlawonline.com/api/Product/CitationPDFURL?file=Journals\GPLR\GPLR2025022.pdf> accessed 2 December 2025.

[36] Sciarrone Alibrandi, Rabitti and Schneider (n 7) 10.

The CCD is successful in addressing some interpretative hurdles that were present in analogous provisions of GDPR[37]. Only recently did the CJEU clarify, in a judgment incidentally related to credit scoring ("SCHUFA case"), that GDPR's art. 22, as far as it concerned a "decision based solely on automatic processing", shall be interpreted broadly[38]. It is thus sufficient for the aforementioned provision to be applicable if the credit scoring is provided by an external agent is then provided to the decision-maker and that decision-maker "draws strongly" on those obtained probabilities to "to establish, implement or terminate a contractual relationship with that person". CCD is much clearer in this respect, as it sets the bar for its applicability much lower, as it suffices that "automated data processing" to be "involved" in the credit scoring.

Yet the content of this right remains anything but clear. It is still vague what exactly is the "logic" of the system, as well its "significance" and "effect" on the decision. Such explanation, in the case of credit scoring, can include a multitude of data, e.g.: just an information an AI system was used in the decision procedure; a detailed overview how that system fits into the totality of decision processes for credit scoring in a given company; identification of concrete data that is fed to the system (and that includes not only identifiable features of the explainee – like their income, but might also pertain to a full body of data used for model's training), information how this data in general affects the output (global explanation) or how it does so in the particular case (local explanation), identification of the algorithm used to arrive at the decisions; finally – a detailed description how the model transformed the input data to generate a final output, expressed for example by a set of mathematical formulae. Moreover, some of this data might be infeasible to produce, due to the complex nature of models used and varied fidelity of explainers used.

As far as the "clarity" and "comprehensibility" is concerned, it should be noted some explanations may be understandable only when the addressee already possesses a sufficient level of AI literacy or possesses a grasp of higher mathematics – a completely unrealistic requirement in the case of laypeople. Further layer of complexity is added by the fact that it is possible that the responsibility for model development and deployment is shared among several entities and they might not be informed of how it works outside of their area of responsibility, due to e.g. technical and communication issues, or trade secrecy[39].

[37] Eurpean Union, 'Regulation (EU) 2016/679 of the European Parliament and of the Council of 27 April 2016 on the Protection of Natural Persons with Regard to the Processing of Personal Data and on the Free Movement of Such Data (General Data Protection Regulation)'.

[38] E. Arroyo Amayuelas, *A Third Directive on Consumer Credit*, "European Review of Contract Law" vol. 20 no. 1 (2024), DOI: 10.1515/ercl-2024-2001; Court of Justice of the EU, Judgment of 7 December 2023 in Case C-634/21 (OQ v Land Hessen, SCHUFA Holding – Scoring), ECLI:EU:C:2023:957.

This interplay between trade secrecy and transparency has been clearly illustrated in the case of the COMPAS system used in the United States for recidivism probability prediction, where the proprietary nature of the algorithm was recognized by the courts as sufficient to accept the lack of explainability of the system, whereas separate rationale for a decision exists[40]. In a number of Dutch childcare benefits cases opaque automated risk-assessment tools contributed to discriminatory treatment of families, and the public authorities fully succumbed to algorithmic bias, relying on algorithmic results because "computer said so" and no further insight was available[41]. Following other authors[42], various court cases from across EU can be recalled[43], whereas on the basis of the GDPR the companies were challenged to present e.g. full source codes, algorithms used in their systems, or concrete mathematical formulas used for credit scoring. In general, the need to balance the economic interests of the companies with the data subjects was acknowledged, and thus it was concluded that under the GPPR the data subject does not need to be presented with information allowing them to fully retrace the steps of the algorithm ("recalculate the score"). In practice, this balancing has been interpreted as permitting disclosure of high-level information about the logic of the system (e.g. the categories of data used, the main factors influencing outcomes, the general purpose of the model, and the role of automation in the decision), while excluding disclosure of proprietary elements such as full source code, or exact model parameters. While the relationship between trade secrecy and explainability deserves a paper of its own, herein we take the stance that even if companies presented the explainees with the source codes (cf. art. 74(13) of the AIA that allows market surveillance authorities to access those codes), weights, raw scores obtained through various explanation methodologies, or provided with the concrete training data, it still would be insufficient to present end user with an actionable knowledge[44], and that information would have to be elaborated on to show how it connects to the context of a given case and social reality (see below).

[39] Górski and others (n 21) 6; Derek Leben, 'Explainable AI as Evidence of Fair Decisions' (2023) 14 Frontiers in Psychology 1069426, 1.

[40] Loomis v. Wisconsin, 881 NW2d 749 (Wis 2016) cert denied 137 S Ct 2290 (2017).

[41] Busuioc, Curtin and Almada, 'Reclaiming Transparency' (n 5) 80.

[42] Metikos and Ausloos (n 6) 23–24.

[43] Landsgericht Traunstein [2024] Regional Court Traunstein 6 O 2465/23, Bundesverwaltungsgericht [2023] Federal Administrative Court Austria ECLI:AT:BVWG:2023:W252.2237416.1.00, Bundesverwaltungsgericht [2023] Federal Administartive Court Austria ECLI:AT:BVWG:2023:W256.2234851.1.00 W256 2234851-1.

[44] Madalina Busuioc, Deirdre Curtin and Marco Almada, 'Reclaiming Transparency: Contesting the Logics of Secrecy within the AI Act' (2023) 2 European Law Open 79, 96.

All the issues mentioned in the preceding paragraphs can eventually lead to an explanation that ends as the lowest possible denominator of legal and technical requirements and possibilities[45]. This has been confirmed in the empirical research, whereas the banks tend to focus on simpler solutions that are accurate enough[46] (and we can confirm this anecdotally from personal experience). In such cases, systems are interpretable (for ML experts) and financial institutions focus on that aspect mainly for model improvement and internal development support, not end user[47].

There is a certain similarity between the rights conferred by the AIA (especially art. 86), the CCD, and the GDPR, as well as the other legal documents. In the case of GDPR the scholarly interpretations diverge whether it provides only a very limited "right to be informed" or fully-fledged "right to explanation" – as far as the normative text is concerned, explainability is explicitly mentioned only in recital 71[48], and the normative part, with central role of arts. 22(3)[49],13(2)(f), 14(2)(g), and 15(1)(h)[50], is more technical and seemingly limited in nature. Yet, according to GDPR, similarly to CCD, the data subject has the right to receive at least "a meaningful information about the *logic* involved, as well as the *significance* and the envisaged *consequences* of such processing for the data subject" (emphasis added). Similarity of the language used makes GDPR an important interpretative indicator regarding the CCD's explanation content. In this context, the interpretation set forth in Advocate General's opinion in the SCHUFA case stated that explanations should include both the global component (aggregate weights of the factors taken into account in the all of the decision-making processes) as well as the local ones (reasons for a certain result). Yet, this is not universally acknowledged. In general, courts do not tend to strive for a detailed recollection of how the result was exactly achieved and what feature of explainee were exactly taken into account[51]. For example, the District Court of Appeals of Amsterdam ruled for a global view of the explanations, noting that only the aggregate weights have to be disclosed, i.e. the ones that pertain to the totality of cases, not the individual one, and the similar stance was taken by the Austrian Federal Administrative Court[52].

[45] Simona Demkova, 'The AI Act's Right to Explanation: A Plea for an Integrated Remedy' (*MediaLaws*, 31 October 2024) <https://www.medialaws.eu/the-ai-acts-right-to-explanation-a-plea-for-an-integrated-remedy/> accessed 21 December 2024.

[46] Kuiper and others (n 19); Giovani Valdrighi and others, 'Best Practices for Responsible Machine Learning in Credit Scoring' (arXiv, 30 September 2024) <http://arxiv.org/abs/2409.20536> accessed 10 December 2025.

[47] Kuiper and others (n 19) 6–7.

[48] "[T]he right … to obtain an explanation of the decision …".

[49] Concerned with automated decision-making under the GDPR being permitted provided that, at a minimum, the individual has the right to express their view, challenge the decision, and obtain human review.

[50] Arts. 13(2)(f), 14(2)(g), and 15(1)(h) use substantively identical wording requiring that individuals be informed of the existence of automated decision-making and provided with meaningful information about the logic involved and the significance and envisaged consequences of such processing.

Yet, the presented views are superficially technocentric and do not take into account the actionability of the knowledge passed on to the explainee, the possible discrepancy between the explanation and how the system arrives at the result, or the existing theoretical frameworks that have been built around how humans justify their findings, in scientific, legal, and other endeavours[53]. It focuses on algorithmic aspects of decision-making process, disregarding the fact that it is embedded in a broader context of social reality. In the following chapter we will argue for a broader understanding of an explanations – in the sense of encompassing the wider number of contexts the decision is made in and not being limited to the algorithm used or even the concrete values of weights associated with the decisive factors.

## IV. AGAINST EXPLAINABILITY. TOWARDS A JUSTIFIABLE AI

The main point of this paper is to deplore (technical) explainability in a B2C relationship. The stance "against explainability" does not mean we support the point of view that in various scenarios – including the exemplary credit scoring – the consumers should be devoid of a rationale for a decision. We assume this position to argue for opposite – that explanations, limited to technical details are not sufficient to grant the credit applicant with actionable knowledge. Yet, the XAI language is still so prevalent, that it is often used as the sole reason for keeping up the legal compliance[54].

Empirical results show that the outputs of contemporary solutions for explainability are alien to persons that do not have ML experiences and are aimed at the developers. The XAI methods are diverse, aimed mainly at technical staff to facilitate their understanding and model improvement[55], and enabling understanding for lay-people is a challenging task[56].

[51] Metikos and Ausloos (n 6) 19.

[52] ECLI:NL:GHAMS:2023:793, Gerechtshof Amsterdam, 200.295.742/01; Austrian Federal Administrative Court, Case BVwG - W252 2246581-1/6E; cf. Metikos and Ausloos (n 4) 18.

[53] Tim Miller, 'Explanation in Artificial Intelligence: Insights from the Social Sciences' [2018] arXiv:1706.07269 [cs] <http://arxiv.org/abs/1706.07269> accessed 15 December 2019.

[54] Hofit Wasserman-Rozen and Ran Gilad-Bachrach, 'LOST IN TRANSLATION: THE LIMITS OF EXPLAINABILITY IN A1I' (2024) 42 423.

[55] ibid.

[56] Maria Riveiro and Serge Thill, '"That's (Not) the Output I Expected!" On the Role of End User Expectations in Creating Explanations of AI Systems' (2021) 298 Artificial Intelligence 103507; Marzyeh Ghassemi, Luke Oakden-Rayner and Andrew L Beam, 'The False Hope of Current Approaches to Explainable

Empirical results show that when different types of explanations (LIME, SHAP, Grad-CAM-based) are presented to prospective explainees, they are unsure on how to make sense of them[57]. Or, in other words, whilst the lawyers are used to working with facts, argumental chains, and evidence, the aforecited research presented them with passages of text with some fragments marked by AI as most relevant for its classification decision. Adjustment to such visualization method was not easy and alien to prospective users. Similarly, in the case of tax law, the lawyers found explanations generated by the technical methods to be completely dissimilar to what they used to work with[58]. In practice, end-user-facing explanations have to come with instructions for use to allow the explainee a starting point[59], thus adding another layer of bias.

Different XAI techniques apply to different groups of addressees, sporting different desiderata, background knowledge, or sets of requirements. Especially when tuned for human understanding, different XAI techniques explain varied aspects of the original model, and the aggregates can be source a confusion when applied to the concrete case (e.g. the model in general can attribute 50% of weight to income, yet in a particular case a credit history might have been a deciding factor)[60]. The algorithms used to arrive at the results do not necessarily reflect how humans arrive at – or justify – their decision[61]. In many cases, XAI serve a subsidiary role to human ingenuity. By that we mean the overview given by XAI methods is limited in scope and does not facilitate full insight into how algorithms arrive at the conclusion[62]. It gives an overview of how the conclusion was arrived at, however this does not map directly to causality. This approximation can only subsequently be used by the engineers in their exploration of the system whilst looking for bugs or biases[63].

There is a number of desiderata concerned with the decision-making in the financial sector[64]. The explanations may be need to be elicited for the use of system developers,

---

Artificial Intelligence in Health Care' (2021) 3 The Lancet Digital Health e745; Jürgen Dieber and Sabrina Kirrane, 'Why Model Why? Assessing the Strengths and Limitations of LIME' [2020] arXiv preprint arXiv:2012.00093.

[57] Górski and Ramakrishna (n 23); Dieber and Kirrane (n 56).

[58] Górski and others (n 21).

[59] Dieber and Kirrane (n 56) 9; Górski and others (n 21) 563.

[60] Metikos and Ausloos (n 6) 19–20.

[61] Almada and others (n 11) 2.

[62] Bundesbank (n 34).

[63] Wasserman-Rozen and Gilad-Bachrach (n 54) 429.

[64] Carlos Mougan, Georgios Kanellos and Thomas Gottron, 'Desiderata for Explainable AI in Statistical Production Systems of the European Central Bank' (arXiv, 12 February 2022) <http://arxiv.org/abs/2107.08045> accessed 10 December 2025; Markus Langer and others, 'What Do We

auditors, internal oversight boards, analysts to support them in day-to-day decision-making, clients. Each of those categories needs explanations tailored to their needs, background knowledge, mental capabilities[65]. We submit there is not a single all-size-fits-all explanations that can be presented to those diverse categories. Contemporary explanation generation algorithms use sophisticated mathematics- and AI-based solutions to elicit how a system works when arriving at its conclusion. The need to create effective systems can trump the need to create ones that are explainable, and in systems used internally the ability to provide explanations that provide actionable feedback to model developers to facilitate its continuous improvement prevails[66]. When explanations aimed at developers are presented to end users, informing them, for example, of mathematical values of weights on an aggregate level, those users are presented with something that is not immediately understandable, is not always faithful to the original model[67], and thus can make right to explanation inoperative.

Yet, explanations should not be deplored *a capite ad calcem*. We are already well aware that how we arrive at the decision and how we justify it are largely disparate aspects. Or, in other words, a distinction has to be made between a context of discovery and a context of justification, between an (technically-leaning) explanation and (socially-leaning) justification[68]. When applied to law, and borrowing the conceptual framework from legal realism, one can see this distinction emerging when comparing the factors that affect legal decision-maker's decision and the reasons that are subsequently presented for that decision[69]. Former includes personal biases, or intuition, the latter – pertains to the justifications that can be based on legal sources. We do not expect to be presented with neuron firing patterns that occurred when a judge was making a decision, or – on a higher level – their emotional state or how exactly their mind wandered when deciding a case, yet in many situations this is what XAI offers. Taking a court-like decision-making analogy further: when a specialistic knowledge is needed in a case, a judge resorts to an expert testimony (like a bank to external credit scoring agency). The court in Polish law is not bound by this opinion[70], but even if in practice the courts rarely diverge from those opinions, their

Want from Explainable Artificial Intelligence (XAI)? – A Stakeholder Perspective on XAI and a Conceptual Model Guiding Interdisciplinary XAI Research' (2021) 296 Artificial Intelligence 103473.

[65] Bundesbank (n 34) 5; Balint Gyevnar, Nick Ferguson and Burkhard Schafer, 'Bridging the Transparency Gap: What Can Explainable AI Learn From the AI Act?' [2023] arXiv preprint arXiv:2302.10766 3.

[66] Kuiper and others (n 19) 2.

[67] Freiesleben and König (n 24) 6.

[68] ibid 7; Górski and others (n 21) 3; Metikos and Ausloos (n 6) 20.

[69] Mireille Hildebrandt, 'Privacy as Protection of the Incomputable Self: From Agnostic to Agonistic Machine Learning' (2019) 20 Theoretical Inquiries in Law 83, 32.

[70] Supreme Court, 'Decision of the Supreme Court of September 25, 2024'.

justification can not be limited to simple acknowledgment that an expert has decided in the case. Similar should be expectations in the case of bank using a credit scoring system. In other works it was noted the technical explainers allow only a skilled personnel to "guesstimate" how the algorithm arrives at the decision[71]. Even in the case of interpretable solutions, those usually need a level of skill to facilitate understanding. What is true, however, is the fact the knowledge passed on to the consumer should be the one that allows them to take relevant action.

What is expected from a justification is a convincing recollection of facts of the case and relevant legal and contractual provisions, coupled together with a logical argumentative chain that can be presented to explainee and relevant organs performing a supervisory role (cf. art. 18(8)(c) of the CCD). In those lines, it should be noted the concrete mathematical formulae as well as numeric values of weights are not useful for the end user. The same goes for simply saying that an AI system was involved. For the purpose of art. 22 the European Data Protection Board, endorses the Guidelines mentioning the following elements as necessary in a justification: "the categories of data that have been or will be used in the profiling or decision-making process; why these categories are considered pertinent how any profile used in the automated decision-making process is built, including any statistics used in the analysis; why this profile is relevant to the automated decision-making process; and how it is used for a decision concerning the data subject"[72]. Thus, instead of exact description of algorithm, a more procedural approach is taken, whereas the model's exact inner working are kept black-boxed and the focus is more on how it integrates into a general shape of decision-making procedure. In other words, larger context is acknowledged when an explanation is presented and it does not boil down to details how model transformed the data on its way to generate an output. Technical XAI language should not to be literally lifted and used in a B2C relationship, as this is limiting an user-facing explanation focused on the context of discovery. Clear and actionable explanations should encompass the social, legal, and procedural contexts and not be simply recollection of possibly deceiving model's aggregate weights, formulas that use those weights, or other technical details. This is also supported by the linguistic analysis of the CCD, as the explanation, according to art. 18(8) should include not only the description of underlying "logic", but also the significance and effects of that "logic". In other words, multitude of technical solutions can be used to uncover the inner-working of the system, but it is up to the human operator to present those results to the explainee in a form that allows them to take action.

[71] Chłopecka (n 8) 81.

[72] Metikos and Ausloos (n 6) 20; 'Automated Decision-Making and Profiling | European Data Protection Board' <https://www.edpb.europa.eu/our-work-tools/our-documents/guidelines/automated-decision-making-and-profiling_en> accessed 19 December 2025.

This facet of the right to explanation has also been recognized by other authors (though their interpretations are sometimes labeled as "creative" [73]). Legal philosophers have noted this disparity between explanation and justification in the case of ML systems[74]. When an algorithm is used, the former is only a heuristic showing how a decision was arrived at, not why it was taken[75]. The public authorities awareness of this issue is also rising, with Berlin Data Protection Authority being the first one that requires the justification of the decision when automatic data processing is involved[76].

## V. CONCLUSION

In this paper, we have explored a notion of XAI, giving the credit scoring context as an illustrative example. Even the simplest ML methods need a body of background mathematical knowledge to facilitate understanding, something acknowledged even in the AIA itself (art. 4). Thus, the argumentation we submitted in this papers proceeds as follows: (1) explanation itself is a technical term, arising from the needs of AI developers and answering their requirements; (2) In B2C relationship the consumers need to be provided with an actionable knowledge so that they are in position to understand the system and protect their interests; and (3) presentation with raw results given by XAI methods or simply acknowledgment that AI models were used does not provide the consumer with operational knowledge.

This is because AI models arrive at the conclusion differently than humans when they process information. The description of underlying logics do not always translate well to format that can be comprehensible for all potential addressees. As evidenced by our answer to RQ1, explanations generated by XAI methods lean heavily on the side of the developers and technical users. They are not necessarily clear, not comprehensible for addressee. We contend that technical explanations alone constitute regulatory fiction: they fail to empower consumers. Justifiability – rooted in legal reasoning – is the only path to genuine transparency. For example, the consumers in the context of credit scoring are granted the right the express their point of view (art. 18(8)(c)), but to contest the decision, they need to be provided with necessary body of arguments to challenge them.

[73] Metikos and Ausloos (n 6) 19.
[74] ibid.
[75] M. Hildebrandt, *Privacy as Protection of the Incomputable Self...*, p. 32; Metikos and Ausloos (n 6) 20.
[76] Metikos and Ausloos (n 6) 20.

In this context, a distinction between the context of discovery and the context of justification, or an explanation and justification, should be acknowledged for. However, explanations, in the technical sense, should be treated as uncovering the internals of the context of discovery and after the decision is made, a context of justification should be acknowledged. The discovery is internal, aimed at arriving the result; the justification is external, aimed at giving the reasons for the result, persuasive. The latter should be included when a consumer is confronted with credit-scoring decision to present them with actionable knowledge to facilitate understanding and contestation.

Thus, answering the RQ2, we submit that the right to explanation has to be interpreted broadly even in current legal environment and – recalling RQ3 – the legal justification should serve as a proper standard for the protection of rights. If the explanations delivered to the explainee are confusing, hard to grasp, they offer them no actionable knowledge and thus are prohibitive as far as the recourse is to be considered. This inadequacy of explanations has already been proven in a number of empirical works, including ours. In this paper, we have taken a stance that the justifiability of the ML-based decisions is a quality that allows to operationalize the explanations and make them useful for the end-user.

Such justifications can in principle be produced even by AI-models, like large-language ones (LLMs), with their impressive capabilities of processing textual modalities[77]. They can be fed data regarding a given credit scoring case, as well as the information obtained from low-level explanation-generation systems and produce, seemingly convincing, justification. However, whilst their reasoning capabilities are still a point of concern and contention among the specialists[78], and such models are prone to hallucinations (presenting false information as factually correct), this can be further supported on a computational level with a layer of logical rules pertaining to most common credit scoring use cases and expressing the expert knowledge of how the system works to further support the generation of justification. This is the aim of future research we are currently pursuing.

[77] Łukasz Górski and Shashishekar Ramakrishna, 'Right to Explanation in LLMs: Lessons from EU AI Act and GDPR' [2025] IT Professional.

[78] Konstantine Arkoudas, 'ChatGPT Is No Stochastic Parrot. But It Also Claims That 1 Is Greater than 1' (2023) 36 Philosophy & Technology 54; Emily M Bender and others, 'On the Dangers of Stochastic Parrots: Can Language Models Be Too Big? 🦜', *Proceedings of the 2021 ACM conference on fairness, accountability, and transparency* (2021).